# Enriching Verbal Feedback from Usability Testing: Automatic Linking of Thinking-Aloud Recordings and Stimulus using Eye Tracking and Mouse Data

Supriya Murali, Tina Walber, Christoph Schaefer, Sezen Lim

The think aloud method is an important and commonly used tool for usability optimization. However, analyzing think aloud data could be time consuming. In this paper, we put forth an automatic analysis of verbal protocols and test the link between spoken feedback and the stimulus using eye tracking and mouse tracking. The gained data – user feedback linked to a specific area of the stimulus – could be used to let an expert review the feedback on specific web page elements or to visualize on which parts of the web page the feedback was given. Specifically, we test if participants fixate on or point with the mouse to the content of the webpage that they are verbalizing. During the testing, participants were shown three websites and asked to verbally give their opinion. The verbal responses, along with the eye and cursor movements were recorded. We compared the hit rate, defined as the percentage of verbally mentioned areas of interest (AOIs) that were fixated with gaze or pointed to with the mouse. The results revealed a significantly higher hit rate for the gaze compared to the mouse data. Further investigation revealed that, while the mouse was mostly used passively to scroll, the gaze was often directed towards relevant AOIs, thus establishing a strong association between spoken words and stimuli. Therefore, eye tracking data possibly provides more detailed information and more valuable insights about the verbalizations compared to the mouse data.

## 1 INTRODUCTION

Think aloud protocol is an important usability tool wherein participants verbalize their thoughts as they perform a usability task. It provides rich data to gain insights into an individual's thought process and attention. However, the analysis of the data can be quite time consuming, especially if it has been recorded, for instance, in an unmoderated remote test. Therefore, an automatic analysis would support usability experts and would, as a consequence, lead to a further spread of usability optimization. In this work, we investigate if a correlation between the stimulus, here a web page, and verbalizations could be established by using interaction data from the computer mouse and eye tracking.

Both eye and cursor movements could provide insights into cognitive processes and hence, could represent the correlation between the stimulus and think aloud data. However, one challenge in eye tracking is that it can be expensive and remote testing is more difficult. Mouse tracking, on the other hand, being easily available, nevertheless, shows a lot of variability among individuals. Hence, in our study, we put forth an automatic transcription of think-aloud data and an association of verbal content to specific areas of the web pages used as stimuli, based on the one hand on mouse movements, and on the other hand, on gaze data. We found a correlation between verbal protocols and the stimulus, but importantly, gaze-based correlations were significantly stronger.

## 2  LITERATURE REVIEW

While previous research has shown an association between eye movements and verbalizations, very few studies, have looked at it from a usability testing perspective. For instance, a consumer behavior study [14] found that when participants compared different products, they exhibited alternating fixation patterns between the choices, and that 83% of such alterations were accompanied by explicit statements of comparison. In fact, according to Cooke and Cuddihy [4] eye data and verbalizations match about 80% of the time. Although this does not necessarily allude to a temporal synchrony, Elling and colleagues [5] found that a verbalization might occur before or after fixation on a relevant region. It has also been suggested that the combined use of eye tracking and verbal protocols could provide additional insights into cognitive processes [15].

As mentioned earlier, eye tracking may not be the most feasible method. Although with webcam tracking [11,12] this has changed, an alternate method which is easily available and remote-friendly is mouse tracking. Studies have found that both eye and mouse movements represent visual attention and are determined by the position and relevance of an object on a search page [12]. Some have even suggested that mouse tracking can be used as a substitute for eye tracking since cursor position can be used to predict gaze. For example, one study showed that the cursor and gaze are present in the same area of the visual field at least 75% of the time [2]. In fact, few studies have also alluded to the fact that cursor movements are related to verbalizations [3,7,9]. However, one issue when looking at cursor movements, as pointed out by Navalpakkam et.al. [10], is that there is wide variability in mouse behavior between subjects. For instance, while some users point with the cursor, others keep idle or use it just to scroll.

## 3  METHOD

### 3.1  Technical setup

For the recording of user behavior, Eyevido Lab, a software for conducting web-based eye tracking analyses was utilized. The study was conducted within a dedicated user research lab, using a 21.5" monitor with a display resolution of 1920x1080 pixels. Data was recorded using a wireless mouse, a compact on-camera microphone with a sampling rate of 48 kHz and a Tobii 4C Pro infrared eye tracking device with a sampling rate of 90 Hz. Each gaze-, mouse- and voice-value has a global timestamp with millisecond accuracy to put them in temporal context.

### 3.2  Participants

Participants were recruited through a targeted mailing list that individuals had voluntarily subscribed to. A total of ten individuals between 25 – 35 years (Mean = 28.5), including four males, five females, and one non-binary person, participated in the study. All participants provided written consent for data processing and accepted the terms and conditions.

### 3.3  Procedure

Prior to the study, participants were briefed on the functionality of an eye tracker and were informed that the subsequent web pages were presented as scrollable screenshots and did not have interactive features. The stimuli consisted of three distinct screenshots extracted from the official websites of German cities or municipalities, namely Berchtesgaden,

Bielefeld and Koblenz (Figure 1). The selection of these websites was based on the visually engaging presentation of content and design, encompassing topics such as nearby tourist attractions, current local news, and cultural events. Each topic is complemented by relevant images displayed in various layouts, resulting in visually diverse sections across the websites. The captured screenshots accurately replicated the scrollable layout, display size, and appearance of the actual websites.

Prior to viewing each webpage, participants were instructed to share their thoughts and opinions aloud. To encourage this process, written prompts were provided, including the questions, "What aspects do you find negative?", "What stands out to you as positive?", and "What elements do you find confusing?". These prompts remained consistent for all three websites. Participants then took between 00:50 and 03:04 minutes (mean = 01:31 minutes) while they verbally expressed any thoughts or observations they had regarding the prompted questions or any other aspects on each web page.

In certain instances, some participants sought clarification while viewing a webpage by asking questions such as "Should I begin now?". In response, the study supervisor instructed them to commence expressing their opinions and to highlight anything that caught their attention. Note that the study was conducted in German, but examples in this paper have been translated to English for convenience.

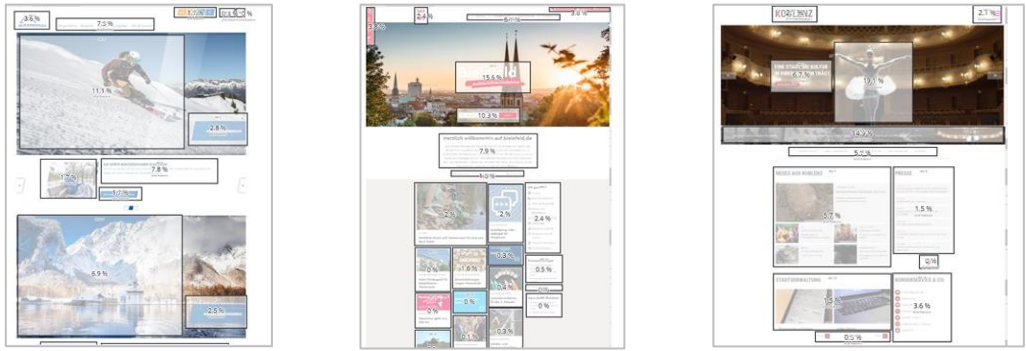

Figure 1: Shows a section of the stimuli, which are the three city websites and the AOIs marked for each one

### 3.4 Marking of AOIs on the stimulus

All relevant AOIs for each page were drawn manually by an expert. They mark general blocks such as the navigation menu, a text section or an image. On average, 42 AOIs were drawn per web page. AOIs could also have been calculated algorithmically, for example, by image processing or analysis of the web page structure. But, at this point, we chose a manual method to avoid any inaccuracies as a result of using an algorithmic method.

### 3.5 Transcription and diarization

Using the combination of two AIs from huggingface [6], we analyzed the audio track of the recorded sessions. With the first AI "whisper" (medium) from OpenAI [13], the spoken words in the screencast were transcribed and time-stamped. A second AI "speaker-diarization" from pyannote [1] was then able to identify and separate the interviewer's and the subject's spoken words. This pipeline consumes few resources (less than 1h on 6 CPUs for the entire dataset) and produces near-perfect results for German voice recordings. The transcripts did not require any manual post-processing.

The speaker-diarization delivered perfect results and distinguished between spoken feedback of the testers and instructions or comments given by the usability expert. In the further analysis, only the testers feedback is used.

## 3.6 Calculation of links between verbal feedback and AOIs

Since timestamps were synchronized for the gaze data, mouse data and spoken sentences, they were used to determine where the subjects looked or pointed and what they said at a specific time. Thus, each spoken sentence was assigned a segment of a gaze path and a mouse path. Using the intersection of the paths with the AOIs, the verbal utterances were assigned to the individual AOIs.

## 4 ANALYSIS

The links between the verbal utterances and AOIs were evaluated to label the hits and calculate the hit rate, which was compared between the gaze and mouse data. Additionally, the sentiments for each statement were marked and analyzed.

### 4.1 Manual labeling of hits and sentiments

Every verbal statement as obtained from the transcription was checked manually to see which AOIs had been mentioned by the subject. There were a total of 344 statements (Mean per subject = 34) made and 231 statements (Mean per subject = 23) that mentioned AOIs. The remaining statements were general ones with no reference to an AOI and were excluded. The mean duration of each statement was 7.2 seconds and 7.4 (SE = 3.3) AOIs were mentioned per statement.

In some instances, a single statement was split into two utterances by the AI transcription. The AOIs were marked separately for them. There were also instances, where the context of a statement had to be checked with the previous statement. For example, subject #8 says, "But I notice that a lot of large images are used." and the next utterance is "Again, I think that's a bit too much.". One would need to check the first statement to understand that the subject is referring to the images in the second statement. Finally, for the sentiment analysis, each statement was manually evaluated and marked with a positive, negative or neutral sentiment. There were a total of 100 positive, 64 negative and 67 neutral statements. The sentiment analysis could also be done based on AI algorithms. But available AIs we could test were unusable for our use case, probably because they were trained on datasets from other domains such as tweets or Amazon reviews whose utterances differ from Think Aloud statements in the usability context. Manual sentiment labeling, on the other hand, was straightforward and excluded AI inaccuracies.

### 4.2 Calculation of hits rate

In order to calculate the hit rate, the following steps were taken:
1. For every statement, the AOIs that had been mentioned were noted (This step has already been explained in the previous section 4.1).
2. From step 1, the total number of mentioned AOIs for every statement was calculated.
3. Count for every statement: how many AOIs were mentioned and simultaneously fixated.
4. Count for every statement: how many AOIs were mentioned and simultaneously pointed by the mouse.
5. Finally, Hit rate = (Number of AOIs calculated in step 3 and 4 divided by the total number of AOIs verbally mentioned in step 2) * 100.

# 5 RESULTS

## 5.1 Association between the verbalizations and AOIs based on the gaze/mouse data

We compared the hit rates, which is defined as the percentage of the verbally mentioned AOIs that are fixated on. A paired-sample T-test revealed a significant difference (T(9) = 4.3, p = .002) between the gaze (mean = 66.2%; SE = 2.1) and mouse (mean = 40%; SE = 5.6). In other words, while the gaze was on an average of 66% of AOIs that the subjects verbally mentioned at a given time, the mouse was only on 40% of them. Figure 2 shows the mean hit rate for the gaze and the mouse.

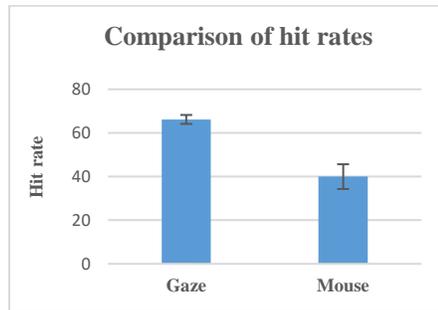

Figure 2: Hit Rate for Gaze vs. Mouse

A closer look at the data revealed that, while the gaze was often directed towards the AOIs that was verbalized, the mouse was used more passively to scroll, which explains the low hit rate for the mouse. Figure 3 shows example screenshots for two different subjects. The gaze path is marked with the numbered circles and the mouse path is shown in the straight-line indicating scrolling. The text below reveals what was said during that time. In both instances, one can observe that,
while the mouse is being used to scroll, the gaze is being directed to relevant AOIs.

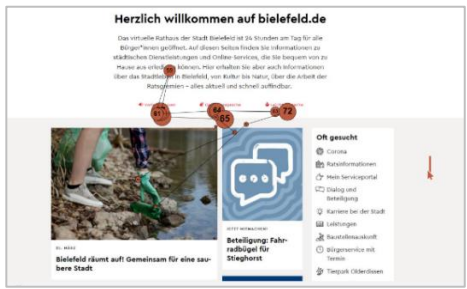 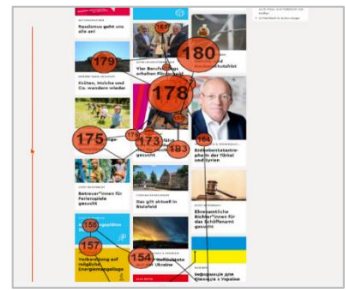

*"What I find interested is that the text can be read aloud, shown in sign language or in easy language."* (Translated from German)

*„Ah, but I don't like that, this blog-like structure of a website, where every topic is interspersed."* (Translated from German)

Figure 3: Screenshots for subjects 8 (left) and 5 (right). The numbered circles show the gaze path and the straight line shows the mouse path. The statement below shows the verbalization during the time-period the screenshot was taken. Note that in several instances, the transcription from the algorithm splits statements. This has been manually corrected for these examples

## 5.2 Is there an effect of sentiments?

There were an average of 10 positive, 6.4 negative and 6.7 neutral statements per subject. Figure 4 below shows the mean hit rate for the positive, negative and neutral sentiments for the gaze and the mouse. A two-factor within subject ANOVA with factors Sentiment (Positive, Negative and Neutral) and Tracking method (Gaze, Mouse) revealed a significant effect of Sentiment ($F(57) = 3.9$, $p = .02$) and Tracking method ($F(57) = 21.5$, $p < .001$), but no significant interaction between the factors ($F(57) = 0.4$, $p = .6$). Specifically, as seen in Figure 4, positive and neutral statements had the highest hit rate, whereas negative statements showed the lowest for both the gaze and the mouse.

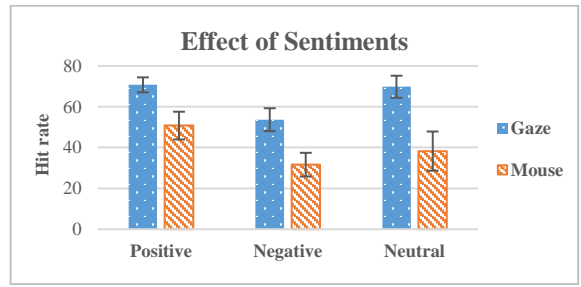

Figure 4: Shows the Hit rate for the three sentiments (Positive, Negative, Neutral) for the gaze (blue, dots) and mouse (red, stripes)

## 6 CONCLUSION AND FUTURE WORK

In our study, we have shown, that a link between spoken feedback and stimulus could be reasonably calculated based on gaze data. 66% of the AOIs of the web page mentioned by the testers were fixated at the same time. On average each statement related to 7 AOIs out of an average 42 marked AOIs per stimulus. The gaze outperforms the mouse data, where only 40% of the AOIs was pointed at. Our findings reveled that gaze-based correlations were significantly stronger. In other words, the gaze was directed more often towards stimuli that were verbally mentioned. The mouse, on the other hand, was mostly used to scroll. Additionally, we also found an effect of sentiments in the association. Importantly, subjects fixated on AOIs they were mentioning more often during positive and neutral assessments. For negative statements, this correlation was the lowest.

Although, previous work has mentioned a correlation between mouse and verbal data [3], it has been pointed out that mouse behavior shows a lot of variability between subjects [10]. Given that eye movements are closely related to attention [8], it is possible that they exhibit more universal patterns and are therefore a better alternative for such automatic analysis. Future studies could focus on further investigation of automatic analysis. Note that, in our study, AOIs and sentiments were manually labelled.

The usage of the generated data - the feedback-stimulus pairs - will also be subject to future research. Possible options include interactive user interfaces for efficiently displaying the feedback to usability experts. Additionally, calculating sentiment maps to highlight areas with positive or negative verbal feedback holds promise in facilitating the analysis of thinking aloud data.